\documentclass[
    reprint,twocolumn,
    aps,prb,10pt,amsmath,amssymb,
    longbibliography,superscriptaddress,
    showpacs,preprintnumbers,
    nofootinbib,
]{revtex4-2}

\usepackage{xr-hyper}
\usepackage[colorlinks=true, linkcolor=blue, citecolor=blue, urlcolor=blue, filecolor=blue]{hyperref}
\usepackage{bm,bbm,bbold}
\usepackage{braket, physics}
\usepackage{mathtools}
\usepackage{amsmath,amssymb, amsfonts}
\usepackage[dvipsnames]{xcolor}
\usepackage{graphicx}
\usepackage[section]{placeins}
\usepackage{multirow}
\usepackage{orcidlink}
\usepackage[normalem]{ulem}  
\usepackage{xspace}
\usepackage{dcolumn}
\usepackage{siunitx}
\usepackage{comment}
\usepackage{float}

\begin{document}

\title{Solving the Peierls-Boltzmann transport equation with matrix product states}

\author{Sangyeop Lee}
\email{sylee@pitt.edu}
\affiliation{Mechanical Engineering and Materials Science, University of Pittsburgh, Pittsburgh, PA 15261, USA}%

\author{Hirad Alipanah}
\affiliation{Mechanical Engineering and Materials Science, University of Pittsburgh, Pittsburgh, PA 15261, USA}%

\author{Juan José Mendoza-Arenas}
\affiliation{Mechanical Engineering and Materials Science, University of Pittsburgh, Pittsburgh, PA 15261, USA}%
\affiliation{Physics and Astronomy, University of Pittsburgh, PA 15261, USA}%

\date{\today}


\begin{abstract}
The Peierls-Boltzmann transport equation (PBE), which governs non-equilibrium phonon transport, suffers from the curse of dimensionality due to its high-dimensional phase space including both real and modal spaces. We explore the use of matrix product states (MPS) for numerical simulation of the PBE. We show that an MPS configuration based on scattering events combined with a dimensionless form of the solution can drastically increase the locality of correlations between tensors in the MPS representation, enhancing its effectiveness in dimension reduction. We further examine the effects of index ordering in an MPS and find that the highest locality is achieved when tensor chains associated with both real and modal spaces are connected from the coarsest grid to each other in the center of the MPS. Using this optimal configuration and a solver inspired by the density matrix renormalization group, we solve the PBE discretized by a finite volume method (FVM). The solution is obtained for crystalline silicon across ballistic, quasi-ballistic, and diffusive transport regimes. An MPS truncated to the compression ratio of $10^{-3}$ suffices to reproduce reference solutions with high fidelity. The computational cost scales sublinearly with the number of grid points in both real and modal spaces, achieving roughly an order of magnitude reduction in computational time compared to the FVM with sparse matrix operation.
\end{abstract}

\maketitle

\section{Introduction}
The Peierls-Boltzmann transport equation (PBE) is the fundamental governing equation describing the non-equilibrium dynamics of phonons in crystalline solids. Despite its broad utility across different applications involving quasi-ballistic phonon transport~\cite{chenNonFourierPhononHeat2021, cahill_nanoscale_2014}, solving the PBE in its original form suffers from the curse of dimensionality~\cite{mazumderBoltzmannTransportEquation2021}. It is an integro-differential equation over a high-dimensional phase space encompassing both real and modal spaces.

Several numerical methods have been developed to circumvent this challenge, each with notable trade-offs. The discrete ordinates method has been widely used, as it reduces the modal-space dimensionality by assuming spherical symmetry in the phonon dispersion and scattering rates~\cite{murthyComputationSubMicronThermal2002,mittalHybridDiscreteOrdinates2011,huGiftBTEEfficientDeterministic2023}. However, even cubic materials such as silicon exhibit appreciable anisotropy in their phonon dispersion relations. The assumption of spherical symmetry prevents the direct use of phonon dispersion and scattering rates obtained from first-principles calculations. This limitation is particularly significant given the success over the past two decades of combining the PBE with \textit{ab initio} inputs to predict bulk thermal conductivity with high accuracy~\cite{lindsay_survey_2018, lindsayPerspectiveInitioPhonon2019}, where the PBE is solved solely in modal space and hence the dimensionality issue is less severe. Monte Carlo methods have been another common approach with considerable success~\cite{peraud_deviational_2014, li_crossover_2019, han_nonequilibrium_2024}. These stochastic methods can readily employ \textit{ab initio} inputs and have proven effective in predicting macroscopic quantities such as temperature and heat flux, which are first-order moments of the distribution function and thus benefit from cancellation of statistical errors across modes. However, Monte Carlo methods suffer from significant statistical noise when computing higher-order quantities such as the local thermal resistivity, which involves the variance of the distribution function around local equilibrium, as well as when performing mode-resolved analyses.

Quantum-inspired algorithms offer a promising route to overcoming the curse of dimensionality, as they have been widely applied to problems requiring high-dimensional solution spaces. In particular, tensor network (TN) methods, originally developed for many-body quantum problems that similarly necessitate representing states in exponentially large Hilbert spaces~\cite{Verstraete_AdvP2008,schollwockDensitymatrixRenormalizationGroup2011,orusPracticalIntroductionTensor2014,orus_tensor_2019,paeckel2019Timeevolutiona,Banuls_2023}, have recently emerged as a powerful tool beyond the quantum realm. In a series of recent studies, TN methods have been applied to solve classical and semi-classical transport equations, including the simulation of classical~\cite{gourianov_quantum-inspired_2022,holscher2025Quantuminspired,gourianov2025Tensor} and quantum~\cite{gomez-lozada_simulating_2025,Connor_2025} turbulence, reacting flows~\cite{adak2024Tensorb,pinkston_matrix_2025}, plasma dynamics~\cite{ye_quantum-inspired_2022,Ye_JPP2024}, flows with nontrivial boundary conditions~\cite{kiffner2023Tensora,peddinti2024Quantuminspired,hulst2025QuantumInspireda}, wave phenomena~\cite{Fraschini_2024, Lively_2025}, and neutron transport~\cite{truong_tensor_2024}.

The key idea underlying TN methods is that, although the solution formally resides in a vast Hilbert space whose dimension grows exponentially with the number of degrees of freedom, the physically relevant solutions often occupy only a small corner of this space~\cite{Verstraete_AdvP2008,orusPracticalIntroductionTensor2014,Bridgeman_2017}. To exploit this structure, one has the freedom to partition the system into subsystems, each designed to hold locality, which will be precised later, and be less dependent on other subsystems. This locality is closely related to the area law of entanglement entropy observed in low-dimensional many-body quantum systems. With such a partitioning, the full high-order solution tensor can be decomposed into a network of smaller low-order tensors connected through bonds. Because inter-subsystem correlations are weak by construction, the bond dimensions can be truncated significantly while incurring only a small and controllable error, resulting in a drastically compressed representation of the solution.

In this paper, we explore the use of a simple one-dimensional TN, namely the matrix product state (MPS)~\cite{Verstraete_AdvP2008,schollwockDensitymatrixRenormalizationGroup2011,Cirac2021rmp,Catarina_EPJB2023}, for solving the PBE. A key to the success of any TN method is identifying a network configuration that maximizes the locality of subsystems and thereby maximally reduces the bond dimensions required for an accurate representation. We therefore begin by examining several different MPS configurations for the PBE in terms of entanglement entropy, systematically comparing indexing schemes for both real and modal spaces. We then solve the PBE using TN-based finite volume method (TNFVM) with the optimal MPS configuration including index ordering identified from this analysis. In our simulation, we consider a one-dimensional real space with the full-dimensional modal space, retaining the complete \textit{ab initio} phonon dispersion and scattering rates. Finally, we examine how the computational cost scales with the dimensionality of the solution space when the TN method is employed. A vast advantage in memory requirements and runtime is found in comparison to conventional finite volume method (FVM), demonstrating the potential of MPS-based approaches for alleviating the curse of dimensionality in the simulations of phonons and other carriers.

\section{Methods}

\begin{figure}[t]
\vspace{0.5cm}
\centering
\includegraphics[width=0.7\linewidth]{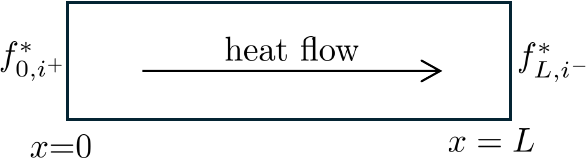}
\caption{\label{fig:domain}
A one-dimensional real-space domain with $f^*_{0,i^+}$ and $f^*_{L,i^-}$, each representing the distribution function $f^*$ on the left and right boundaries as boundary conditions. The subscripts $i^+$ and $i^-$ indicates modes propagating along positive and negative $x$-directions, respectively. 
}
\end{figure}

We consider a one-dimensional real-space domain as shown in Fig.~\ref{fig:domain}. A higher temperature is applied at the left boundary than at the right boundary, so that heat flows in the positive $x$ direction. The temperature difference is assumed to be small and the transport is in linear response regime.

The PBE under the relaxation time approximation is
\begin{equation}
    \label{eq:PBE}
    \Lambda_i \nabla_x f_{x,i}^* = -f_{x,i}^{*} + f_{x,i}^{*0}.
\end{equation}
where the left and righthand sides represent the advection and scattering, respectively. The $\Lambda_i$ is a mean free path (MFP) $v_{i} \tau_i$ of mode $i$ where $v_{i}$ and $\tau_i$ are the $x$-component of the group velocity vector and the relaxation time, respectively.
 The $f_{x,i}^{*}$ is a dimensionless form of phonon distribution function $f_{x,i}$ of mode $i$ at position $x$, defined as
\begin{equation}
    \label{eq:f*}
    f_{x,i}^{*}=\frac{f_{x,i}-f_{i}^{\mathrm{eq}}(T_0)}{f_{i}^{\mathrm{eq}}(T(0))-f_{i}^{\mathrm{eq}}(T(L))},
\end{equation}
where $f^{\mathrm{eq}}$ is the equilibrium Bose--Einstein distribution and $T_0=(T(0)+T(L))/2$. The quantity $f_{x,i}^{*0}$ in Eq.~\eqref{eq:PBE} is the dimensionless local equilibrium distribution, obtained by replacing $f_{x,i}$ in Eq.~(\ref{eq:f*}) with the local equilibrium distribution $f_{x,i}^0$. The local equilibrium distribution $f_{x,i}^{0}$ can be determined from the energy conservation upon scattering:
\begin{equation}
    \label{eq:T0}
    \sum_{i=1}^{n_i} \frac{\omega_i}{\tau_i}f_{x,i}^{0} = \sum_{i=1}^{n_i} \frac{\omega_i}{\tau_i}f_{x,i},
\end{equation}
where $\omega_i$ is the phonon frequency of mode $i$ and $n_i$ is the total number of modes. Hence, Eq.~\eqref{eq:PBE} can be written as
\begin{equation}
    \label{eq:PBE2}
    \Lambda_i \nabla_x f_{x,i}^* = -f_{x,i}^* + \Gamma^{-1} \sum_{j=1}^{n_i} \gamma_j f_{x,j}^*,
\end{equation}
where the coefficients $\gamma_i$ and $\Gamma$ are defined as
\begin{equation}
\label{eq:gamma}
\gamma_i = \frac{\omega_i}{\tau_i} \eval{\frac{\partial f_i^{\mathrm{eq}}}{\partial T}}_{T=T_0},
\end{equation}
\begin{equation}
\label{eq:Gamma}
\Gamma = \sum_{i=1}^{n_i} \gamma_i.    
\end{equation}
Assuming the linear response regime, $f_{x,i}^{*0}$ is the same as dimensionless deviational temperature $T^*$ defined as ${(T(x)-T_0)}/{(T(0)-T(L))}$. It is noteworthy that solving for dimensionless distribution $f^*$, instead of $f$, is critically important to increase the locality in the MPS representation, as will be discussed later.
\subsection{Finite Volume Method}

Applying the FVM with the first-order upwind scheme to the differential operator in Eq.~(\ref{eq:PBE2}), the discretized equations are written as follows. For mode $i^+$ with $v_{i^+} > 0$,
\begin{equation}
    \label{eq:FVM+}
    \begin{split}
        \Lambda_{i^+}^* f_{x,i^+}^* &- (1-\delta_{x,1})\Lambda_{i^+}^* f_{x-1,i^+}^* \\
        &+ f_{x,i^+}^* - \Gamma^{-1} \sum_{j=1}^{n_i} \gamma_j f_{x,j}^* = \delta_{x,1} \Lambda_{i^+}^* f^*_{0,i^+},
    \end{split}
\end{equation}
and for mode $i^-$ with $v_{i^-} < 0$,
\begin{equation}
    \label{eq:FVM-}
    \begin{split}
        -\Lambda_{i^-}^* f_{x,i^-}^* &+ (1 - \delta_{x,n_x})\Lambda_{i^-}^* f_{x+1,i^-}^* \\
        &+ f_{x,i^-}^* - \Gamma^{-1} \sum_{j=1}^{n_i} \gamma_j f_{x,j}^* = -\delta_{x,n_x} \Lambda_{i^-}^* f^*_{n_x+1,i^-}.
    \end{split}
\end{equation}
The subscript $x$ denotes a control volume index ranging from 0 (left boundary) to $n_x+1$ (right boundary) and $n_x$ is the total number of control volumes in $x$-domain. On the left-hand side of Eqs.~(\ref{eq:FVM+}) and~(\ref{eq:FVM-}), the first two terms represent advection and the remaining two terms represent scattering. The quantity $\Lambda_i^*$ is a dimensionless MFP defined as
\begin{equation}
\label{eq:Lambda}
\Lambda_i^* = \frac{v_i \tau_i} {\Delta x},
\end{equation}
where $\Delta x$ is the control volume width. Note that the Dirichlet boundary condition is shown on the righthand side in Eq.~\eqref{eq:FVM+} and Eq.~\eqref{eq:FVM-}.

Equations~(\ref{eq:FVM+}) and (\ref{eq:FVM-}) can be written compactly as
\begin{equation}
    \label{eq:FVMmatrix}
    \mathbf{H} \mathbf{f}^* = \mathbf{s},
\end{equation}
where $\mathbf{s}$ contains the source terms arising from the boundary conditions. The matrix $\mathbf{H}$ is prohibitively large, as it encompasses both the differential operation in real space and the integral operation in modal space. Equation~(\ref{eq:FVMmatrix}) is therefore iteratively solved by separating the differential and integral operations. For mode $i^+$,
\begin{equation}
    \label{eq:FVMiter+}
    \begin{split}
        &(\Lambda_{i^+}^* + 1) f_{x,i^+}^{*,(m)} + (1 - \delta_{x,1})(-\Lambda_{i^+}^*) f_{x-1,i^+}^{*,(m)} = \\
        &\Gamma^{-1} \sum_{i=1}^{n_i} \gamma_i f_{x,i}^{*,(m-1)} + \delta_{x,1} \Lambda_{i^+}^* f_{0,i^+}^*,
    \end{split}
\end{equation}
and for mode $i^-$,
\begin{equation}
    \label{eq:FVMiter-}
    \begin{split}
        &(-\Lambda_{i^-}^* + 1) f_{x,i^-}^{*,(m)} + (1 - \delta_{x,n_x})(\Lambda_{i^-}^*) f_{x+1,i^-}^{*,(m)} = \\
        &\Gamma^{-1} \sum_{i=1}^{n_i} \gamma_i f_{x,i}^{*,(m-1)} -\delta_{x,n_x} \Lambda_{i^-}^* f_{n_x+1,i^-}^*,
    \end{split}
\end{equation}
where the superscript $(m)$ denotes the current iteration step. For each mode, collecting the spatial cells $x = 1,\dots,n_x$ into a vector $\mathbf{f}_{i^\pm}^{*,(m)} = [f_{1,i^\pm}^{*,(m)}, \dots, f_{n_x,i^\pm}^{*,(m)}]^\top$, the Eq.~\eqref{eq:FVMiter+} and Eq.~\eqref{eq:FVMiter-} take the compact form
\begin{equation}
    \label{eq:FVMitercompact}
    \mathbf{H}_{\mathrm{x},i^+}\,\mathbf{f}_{i^+}^{*,(m)} = \mathbf{b}_{i^+}^{(m-1)}, \qquad
    \mathbf{H}_{\mathrm{x},i^-}\,\mathbf{f}_{i^-}^{*,(m)} = \mathbf{b}_{i^-}^{(m-1)},
\end{equation}
with bidiagonal coefficient matrices
\begin{equation}
    \mathbf{H}_{\mathrm{x},i^+} =
    \begin{bmatrix}
        \Lambda_{i^+}^*+1 & & & \\
        -\Lambda_{i^+}^* & \Lambda_{i^+}^*+1 & & \\
        & \ddots & \ddots & \\
        & & -\Lambda_{i^+}^* & \Lambda_{i^+}^*+1
    \end{bmatrix},
\end{equation}
\begin{equation}
    \mathbf{H}_{\mathrm{x},i^-} =
    \begin{bmatrix}
        -\Lambda_{i^-}^*+1 & \Lambda_{i^-}^* & & \\
        & -\Lambda_{i^-}^*+1 & \Lambda_{i^-}^* & \\
        & & \ddots & \ddots \\
        & & & -\Lambda_{i^-}^*+1
    \end{bmatrix},
\end{equation}
which are lower- and upper-bidiagonal for modes $i^+$ and $i^-$, respectively, and right-hand sides
\begin{equation}
    \mathbf{b}_{i^\pm}^{(m-1)} = \Gamma^{-1}\sum_{i=1}^{n_i}\gamma_i\,\mathbf{f}_{i}^{*,(m-1)} + \mathbf{s}_{i^\pm},
\end{equation}
where the boundary vector $\mathbf{s}_{i^\pm}$ has a single nonzero entry,
\begin{equation}
    \mathbf{s}_{i^+} = \Lambda_{i^+}^* f_{0,i^+}^*\,\mathbf{e}_1, \qquad
    \mathbf{s}_{i^-} = -\Lambda_{i^-}^* f_{n_x+1,i^-}^*\,\mathbf{e}_{n_x},
\end{equation}
with
\begin{equation}
    \mathbf{e}_1 = [1, 0, \dots, 0]^\top, \qquad
    \mathbf{e}_{n_x} = [0, \dots, 0, 1]^\top.
\end{equation}

We take the FVM solution converged with a stringent criteria as the reference solution, $\mathbf{f}^*_{\mathrm{ref}}$. Convergence is assessed using spectral radius estimation\cite{ferziger2002computational}. The error at the $m$-th iteration is estimated as
\begin{equation}
    \epsilon \approx \frac{\left\|\mathbf{f}^{*,(m)} - \mathbf{f}^{*,(m-1)}\right\|}{1 - r},
\end{equation}
where the spectral radius $r$ is estimated by
\begin{equation}
    r \approx \frac{\left\|\mathbf{f}^{*,(m)} - \mathbf{f}^{*,(m-1)}\right\|}{\left\|\mathbf{f}^{*,(m-1)} - \mathbf{f}^{*,(m-2)}\right\|}.
\end{equation}
The $\norm{\cdot}$ denotes the $L^2$ norm. The reference solution is considered converged when $\epsilon < 10^{-12}$. 

We consider crystalline silicon at 300~K, with phonon dispersion relations and scattering rates calculated from first principles. The sample length $L$ varies from 1~nm to 1~$\mu$m, spanning the ballistic to quasi-ballistic transport regimes. The boundary conditions for the finite-$L$ cases are $f^{*}_{0,i^+} = \tfrac{1}{2}$ and $f^{*}_{0,i^-} = -\tfrac{1}{2}$. The $L = 1~\mu$m case yields an effective thermal conductivity of 89.1~W/m$\cdot$K, substantially lower than the bulk value of 132.5~W/m$\cdot$K, indicating significant classical size effects. We also consider the diffusive transport limit ($L \to \infty$) using a 1~$\mu$m sample with the known analytical solution of the PBE for this regime imposed as boundary conditions:
\begin{equation}
    \label{eq:fbulk_L}
    f^{*,\mathrm{bulk}}_{0,i^+}=\frac{1}{2}+\frac{(\tau v)_{i^+}}{L},
\end{equation}
\begin{equation}
    \label{eq:fbulk_R}
    f^{*,\mathrm{bulk}}_{n_x+1,i^-}=-\frac{1}{2}+\frac{(\tau v)_{i^-}}{L}.
\end{equation}
Equation~\ref{eq:FVMitercompact} is solved using LU decomposition provided by the LinearAlgebra package in Julia programming language. 

\subsection{Tensor Network Finite Volume Method} \label{sec_TNFVM}

\begingroup
To leverage TN algorithms, all the arrays or functions such as $f_{x,i}^*$, $\Lambda^*_i$, and $\gamma_i$, are encoded in a quantum state of a composite qudit system. For simplicity, we first restrict the discussion to qubits. As an example, we briefly illustrate the encoding process of the distribution function $f_{x,i}^*$ in Fig.~\ref{fig:MPSdecomposition} using tensor diagrams. Fig.~\ref{fig:MPSdecomposition}(a) shows the distribution function $f^*_{x,i}$ as an order-two tensor of size $n_x \times n_i$, where each thick leg represents one of the two indices $x$ and $i$, with dimensions $n_x$ and $n_i$, respectively.

\begin{figure}[h]
    \vspace{0.1cm}
    \centering
    \includegraphics[width=0.8\linewidth]{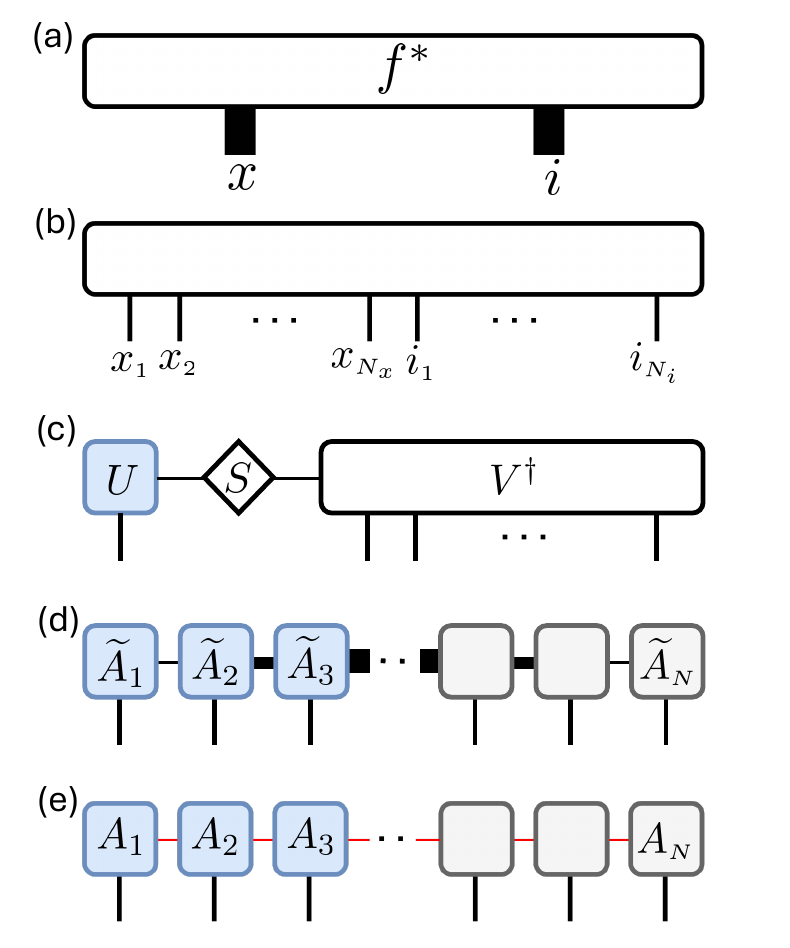}
    \caption{\label{fig:MPSdecomposition}
    Encoding of the distribution function $f^*_{x,i}$ into an MPS. (a) The original distribution as an order-two tensor, with thick legs representing the high-dimensional indices $x$ (dimension $n_x$) and $i$ (dimension $n_i$). (b) The tensor after reshaping indices into binary form, with $N_x + N_i$ legs of dimension two. (c) The result of a single SVD applied to the reshaped tensor. (d) The fully decomposed MPS obtained by sequential SVDs without truncation; thick bonds indicate that the bond dimension grows exponentially toward the center. (e) The compact MPS obtained when the SVDs are performed with truncation; truncated bonds are shown in red.}
\end{figure}

The encoding starts by treating each phase-space point $(x,i)$ as a basis state of a composite quantum system. The indices $(x,i)$ are expressed in binary form $(x_1, \ldots, x_{N_x}, i_1, \ldots, i_{N_i})$, where each bit takes the value $0$ or $1$ and the number of qubits required is $N_\alpha = \lceil \log_d n_\alpha \rceil$ with $\alpha \in \{x,i\}$ and $d=2$ for qubits. After this reshaping, the order-two tensor in Fig.~\ref{fig:MPSdecomposition}(a) becomes the high-order tensor
\begin{equation}
    f^*_{x_1, \ldots, x_{N_x},\, i_1, \ldots, i_{N_i}} = f^*_{x,i},
\end{equation}
shown in Fig.~\ref{fig:MPSdecomposition}(b), in which each of the $N_x + N_i$ thin legs has dimension $2$.

This high-order tensor can be decomposed into an MPS of $N = N_x + N_i$ tensors using, for example, sequential singular value decompositions (SVDs). Figure~\ref{fig:MPSdecomposition}(c) shows the tensor after the first SVD step:
\begin{equation}
    f^*_{x_1, \ldots} = \sum_{\alpha_1} U_{x_1,\alpha_1}S_{\alpha_1,\alpha_1}V^\dagger_{\alpha_1,x_2, \ldots ,i_{N_i}},
\end{equation}
where $\alpha_1$ is an index for the shared bonds by $U$, $S$, and $V^\dagger$. Then, the singular value matrix $S$ is contracted into $V^\dagger$, so that the remaining left block is left-canonical. Repeating the SVD on the right block produces the fully decomposed MPS in Fig.~\ref{fig:MPSdecomposition}(d). In the quantum-state picture, $f^*_{x_1, \ldots, x_{N_x}, i_1, \ldots, i_{N_i}}$ is the amplitude of the basis state $\ket{x_1, \ldots, x_{N_x}, i_1, \ldots, i_{N_i}}$ in the expansion of $\ket{f^*}$, and the MPS represents this state as a chain of $N$ tensors, $\widetilde{A}_{1} \cdots \widetilde{A}_{l} \cdots \widetilde{A}_{N}$, where the first $N_x$ tensors and the remaining $N_i$ tensors are associated with real and modal spaces, respectively. Neighboring tensors $\widetilde{A}_{l}$ and $\widetilde{A}_{l+1}$ share bond $l$ whose dimension determines the amount of quantum correlations (entanglement) between the first $l$ qubits and the remaining $N-l$ qubits.

Without truncation, the MPS of Fig.~\ref{fig:MPSdecomposition}(d) offers no compression over the original tensor of Fig.~\ref{fig:MPSdecomposition}(a). The bond $l$ has dimension $2^{\min(l,\, N-l)}$, growing exponentially toward the center of the chain as shown with thick bonds in Fig.~\ref{fig:MPSdecomposition}(d). Compression is achieved by truncating each bond during the decomposition, retaining only the largest $\chi_l$ singular values and discarding the rest. Fig.~\ref{fig:MPSdecomposition}(e) shows the resulting compact MPS, ${A}_{1} \cdots {A}_{l} \cdots {A}_{N}$, with truncated bonds drawn in red.

Discarding singular values introduces an error. For a tensor of which singular values are normalized such that the sum of the squared singular values is 1, the truncation error at bond $l$ upon retaining the $\chi_l$ largest singular values is
\begin{equation}
    \label{eq:epsilon_l}
    \epsilon_l(\chi_l) = \sum_{j=\chi_l + 1}^{\chi_l^0} \lambda_{l,j}^2,
\end{equation}
where $\lambda_{l,j}$ is the $j$-th largest singular value at bond $l$ and $\chi_l^0$ is the total number of singular values at that bond. We note that, in the context of a bipartition of the quantum state at bond $l$, these singular values coincide with the Schmidt coefficients of the bipartition and are therefore also referred to as Schmidt values. The error in Eq.~\eqref{eq:epsilon_l} is small whenever the Schmidt spectrum decays quickly. In the extreme case of a single nonzero singular value with all other singular values being zero, $\epsilon_l$ vanishes for any $\chi_l \geq 1$. In another extreme case where all $\chi_l^0$ singular values are equal, no truncation can be performed without causing significant error.

The decay of the Schmidt spectrum at each bond can be quantified by the von Neumann entanglement entropy,
\begin{equation}
    \label{eq:entanglement_entropy}
    S_l = -\sum_{j=1}^{\chi_l^0} \lambda_{l,j}^2 \ln \lambda_{l,j}^2,
\end{equation}
which was originally introduced to characterize the uncertainty in measurements of a quantum subsystem due to entanglement with its complement. In the present context, $S_l$ measures the degree of quantum correlation between the sets of qubits to the left and to the right of bond $l$ when the full tensor is bipartitioned at that bond. The entropy attains its minimum value, $S_l = 0$, when a single singular value is nonzero and all other singular values are zero, indicating that the bond carries no correlation. The entropy attains its maximum value, $S_l = \ln \chi_l^0$, when all singular values are equal. Consequently, a bond with low entanglement entropy can be truncated aggressively with negligible loss of accuracy, and the faster the Schmidt spectrum decays, the lower the entropy and the more compact the resulting MPS representation. We refer to this property, fast singular-value decay across the MPS bonds, as a high degree of locality of the MPS representation.

The matrix operator $\mathbf{H}$ can likewise be encoded as a matrix product operator (MPO) \cite{schollwockDensitymatrixRenormalizationGroup2011}. In this MPO--MPS representation, Eq.~(\ref{eq:FVMmatrix}) becomes
\begin{equation}
    \label{eq:TNFVM}
    H \ket{f^*} = \ket{s},
\end{equation}
where the total Hamiltonian $H$ is expressed as
\begin{equation}
    \label{eq:H}
    H = H_{\mathrm{adv}} + H_{\mathrm{scatt}}.    
\end{equation}
The advection operator $H_\mathrm{adv}$ can be expressed as
\begin{equation}
    H_\mathrm{adv}=D^+ \otimes \Lambda^{*,+} + D^- \otimes \Lambda^{*,-}, 
\end{equation}
where $D$ and $\Lambda$ are first-order differential operator based on up-wind scheme and the MFP operator, respectively. The superscripts $+$ and $-$ denote the modes $i^+$ and $i^-$ propagating along $+x$ and $-x$ directions, respectively. The MFP operators, $\Lambda^{*,+}$ and $\Lambda^{*,-}$ can be obtained by encoding the functions $(\tau v)_{i^+} / \Delta x$ and $(\tau v)_{i^-} / \Delta x$ to the MPS through sequential SVDs as described in Fig.~\ref{fig:MPSdecomposition} and no truncation was performed during this process. In order to convert the MPS to MPO, we replaced each leg with order-three $\delta-$tensor, $\delta_{i,j,k}$, which is 1 when all indices $i,j,k$ are the same and 0 otherwise \cite{biamonteTensorNetworksNutshell2017}. The differential operator can be expressed as
\begin{equation}
    D^+ = \frac{I_x - S^-_x}{\Delta x}
\end{equation}

\begin{equation}
    D^- = \frac{S^+_x - I_x}{\Delta x}
\end{equation}
where $S^-_x$ and $S^+_x$ are shift operators to the $-x$ and $+x$ directions. The shift operators can be constructed from scratch in an MPO structures as described in detail in other literature~\cite{ye_quantum-inspired_2022,gomez-lozada_simulating_2025,pinkston_matrix_2025}. The $I_x$ is an identity operator acting on real space.

The scattering Hamiltonian, $H_\mathrm{scatt}$, can be obtained from the last two terms on the lefthand side of Eq.~\eqref{eq:FVM+} and Eq.~\eqref{eq:FVM-} and is expressed as:
\begin{equation}
    \label{eq:Hscatt}
    H_\mathrm{scatt}=I_x \otimes (I_i - \Gamma^{-1}\ket{+}^{\otimes N_i}\bra{\gamma})
\end{equation}
where $I_i$ is an identity operator acting on the modal space. The $\bra{\gamma}$ is obtained by encoding the corresponding function in Eq.~\eqref{eq:gamma} into the MPS form through sequential SVDs as described in Fig.~\ref{fig:MPSdecomposition}. The $\Gamma$ is a coefficient defined in Eq.~\eqref{eq:Gamma}. The state $\ket{+}^{\otimes N_i}$ represents a uniform superposition of all basis states in the modal space.
\endgroup

\begin{figure}[t]
\vspace{0.1cm}
\centering
\includegraphics[width=0.8\linewidth]{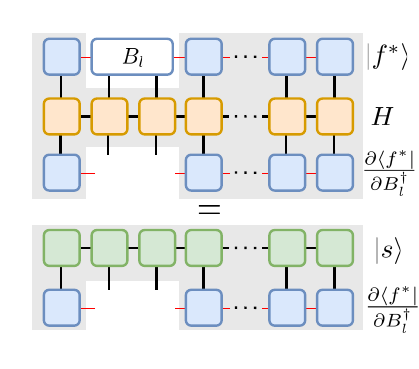}
\caption{\label{fig:DMRG-like}
Tensor diagram of the DMRG-like method for solving Eq.~(\ref{eq:TNFVM}). The local tensor block $B_l$ is optimized while all other tensors in the MPS are held fixed. The shaded area on the top and the bottom represent $H_{\mathrm{eff},l}$ and $s_{\mathrm{eff},l}$ from Eq.~\eqref{eq:DMRG}, respectively.}
\end{figure}

We solve Eq.~\eqref{eq:TNFVM} using a method similar to the two-site density
matrix renormalization group (DMRG), as illustrated schematically in
Fig.~\ref{fig:DMRG-like}. Rather than determining the entire solution tensor
$f^*_{x,i}$ at once, the DMRG-like method optimizes a two-site block at a time.
A two-site block $B_l$ is formed by contracting neighboring tensors $A_l$ and
$A_{l+1}$, and Eq.~\eqref{eq:TNFVM} is projected onto the environment tensors,
namely all tensors of $f^*$ except the two being optimized. This projection is
equivalent to applying the variational principle to the functional
$\bra{f^*}H\ket{f^*} - \bra{f^*}\ket{s}$ with respect to the local block
$(B_l)^{\dagger}$, as shown in Fig.~\ref{fig:DMRG-like}. This yields the reduced
equation

\begin{equation}
    \label{eq:DMRG}
    H_{\mathrm{eff},l} B_l = s_{\mathrm{eff},l},
\end{equation}
where $H_{\mathrm{eff},l}$ and $s_{\mathrm{eff},l}$ are the effective
Hamiltonian and source vector obtained by contracting all tensors except $B_l$.
Their dimensions are $(\chi^2 d^2)^2$ and $\chi^2 d^2$, respectively, where
$\chi$ is the bond dimension connecting neighboring tensors and $d$ is the
dimension of the physical site leg. These dimensions are much smaller than
those of the full structures $H$ and $\ket{s}$, making the local problem
tractable. We solve Eq.~\eqref{eq:DMRG} using the generalized minimal residual
(GMRES) algorithm~\cite{saad1986gmres} in Krylov subspace. After solving, $B_l$
is decomposed back into two tensors, $A_l$ and $A_{l+1}$, by an SVD, and the
bond dimension $\chi$ is set by truncating the singular values according to a
prescribed tolerance. The algorithm then advances to the next block $B_{l+1}$,
formed from $A_{l+1}$ and $A_{l+2}$. This process sweeps through the entire MPS
chain from the left end to the right and back, and the sweep is repeated until
convergence is
achieved~\cite{schollwockDensitymatrixRenormalizationGroup2011,Catarina_EPJB2023}.

For both the FVM and TNFVM, the initial guess is a null array. In the
latter, this corresponds to an MPS in which all elements are zero. From this
initial state, 2--13 sweeps are sufficient to reach convergence, depending on
the specifics of the calculation such as bond dimension $\chi$ and the number of tensors $N$ in the MPS. The convergence criteria is that a relative error $\lVert \mathbf{f}^* - \mathbf{f}^*_{\text{ref}} \rVert / \lVert \mathbf{f}^*_{\text{ref}} \rVert $ is less than $10^{-2}$. The $\mathbf{f}^*$ is the distribution function obtained by TNFVM, and the $\mathbf{f}^*_{\text{ref}}$ is the reference distribution function obtained by FVM with $\epsilon < 10^{-12}$. All MPS and MPO operations are implemented using the ITensors package~\cite{fishmanITensorSoftwareLibrary2022} in the Julia programming language.

\section{Results and Discussion}

\subsection{MPS configuration with minimum entanglement}

The locality of correlations in an MPS depends strongly on how indices are ordered in real and modal spaces. In the MPS of Fig.~\ref{fig:MPSdecomposition}(e), the indices are expressed in binary. Taking the convention that the most significant bit lies on the rightmost site and the least significant bit on the leftmost site, the tensors $A_{N_x}$ and $A_{N_x+N_i}$ carry the information distinguishing $f^*$ at the coarsest scale of index space, whereas $A_{1}$ and $A_{N_x+1}$ carry the information distinguishing $f^*$ at the finest scale. Because the assignment of indices to grid points in real and modal spaces is not unique, the chosen indexing scheme determines both the arrangement of tensors in the MPS and the resulting degree of locality.

{For the real space, we follow the multigrid indexing scheme~\cite{lubaschMultigridRenormalization2018,garcia-ripollQuantuminspiredAlgorithmsMultivariate2021} which has been commonly used for solving transport equations with TN~\cite{gourianov_quantum-inspired_2022, ye_quantum-inspired_2022, pinkston_matrix_2025, gomez-lozada_simulating_2025}. In this scheme, the $x$-grid points are arranged so that the index value increases with the $x$-coordinate. In the binary representation, the most significant digit corresponds to the coarsest grid of the real space domain, effectively distinguishing the left and right halves of the domain. The least significant digit corresponds to the finest grid level, resolving neighboring control volumes. Thus, each of the first $N_x$ tensors encodes the information associated with a different resolved length scale of the spatial domain. 

Since the distribution function of a given mode varies slowly within a small region of real space particularly within length scales shorter than the MFP, the hierarchical structure of the multigrid scheme is expected to promote locality of correlations between qubits encoding similar length scales. For smooth or slowly varying distributions in real space, qubits representing small length scales are not critically necessary to construct the correct distribution function. This is translated to rapidly decaying Schmidt spectra in the bond connecting the qubits representing small length scales to those representing larger length scales. In the extreme case where the distribution function is exactly constant in real space, all bonds have only one non-zero singular value with all other singular values being zero. In the opposite extreme case where the distribution function is randomly fluctuating in all length scales, all qubits are critically required to construct the correct distribution function, making all singular values in each bond non-negligible.
}

Our primary focus in this work is the indexing scheme for the modal space. We examine three different schemes: indexing modes by frequency, by MFP, and randomly. Frequency-based indexing is a common choice for one-dimensional representations of modal space, as it groups modes with similar frequencies and thus similar equilibrium distributions and scattering rates. However, from a simple gas kinetic theory picture, the distribution at a given spatial location is expected to depend more directly on the MFP than on the frequency. Modes with similar MFP at a given location $x$ exhibit similar temperature as their previous scattering process occurred in a similar location at $x-\tau_i v_i$. When the distribution is cast in the dimensionless form of Eq.~(\ref{eq:f*}), this similarity is preserved across modes spanning a wide range of the phonon spectrum, because the normalization removes the strong frequency dependence present in the regular distribution $f$. As a result, the dimensionless distributions of modes with similar MFP closely resemble one another even if their frequencies differ substantially, thus $f^*$ becomes a smooth slowly-varying function in the modal space. Therefore, indexing by MFP, combined with the dimensionless form of the distribution, is expected to maximize locality of correlations in an MPS. The last scheme, random indexing, serves as a worst-case baseline for comparison.

To determine which indexing scheme allows for optimal MPS representation, we
analyze measures of correlation in reference solutions of the PBE obtained
using FVM, $\mathbf{f}^*_{\text{ref}}$. For this, we encode the reference
solutions into MPS under each of the three modal-space indexing schemes.
This encoding is performed by sequential SVDs
~\cite{schollwockDensitymatrixRenormalizationGroup2011,Catarina_EPJB2023} as described in the Method section. We then examine the entanglement entropy at each bond $l$ using Eq.~\eqref{eq:entanglement_entropy}. A low entanglement entropy indicates that the two subsystems are only weakly correlated, implying that the bond dimension can be truncated aggressively
with minimal loss of accuracy. Therefore, the indexing scheme that yields
the lowest entanglement entropy across bonds is expected to make the most
compact MPS representation.

\begin{figure}[t]
    \vspace{0.1cm}
    \centering
    \includegraphics[width=1.0\linewidth]{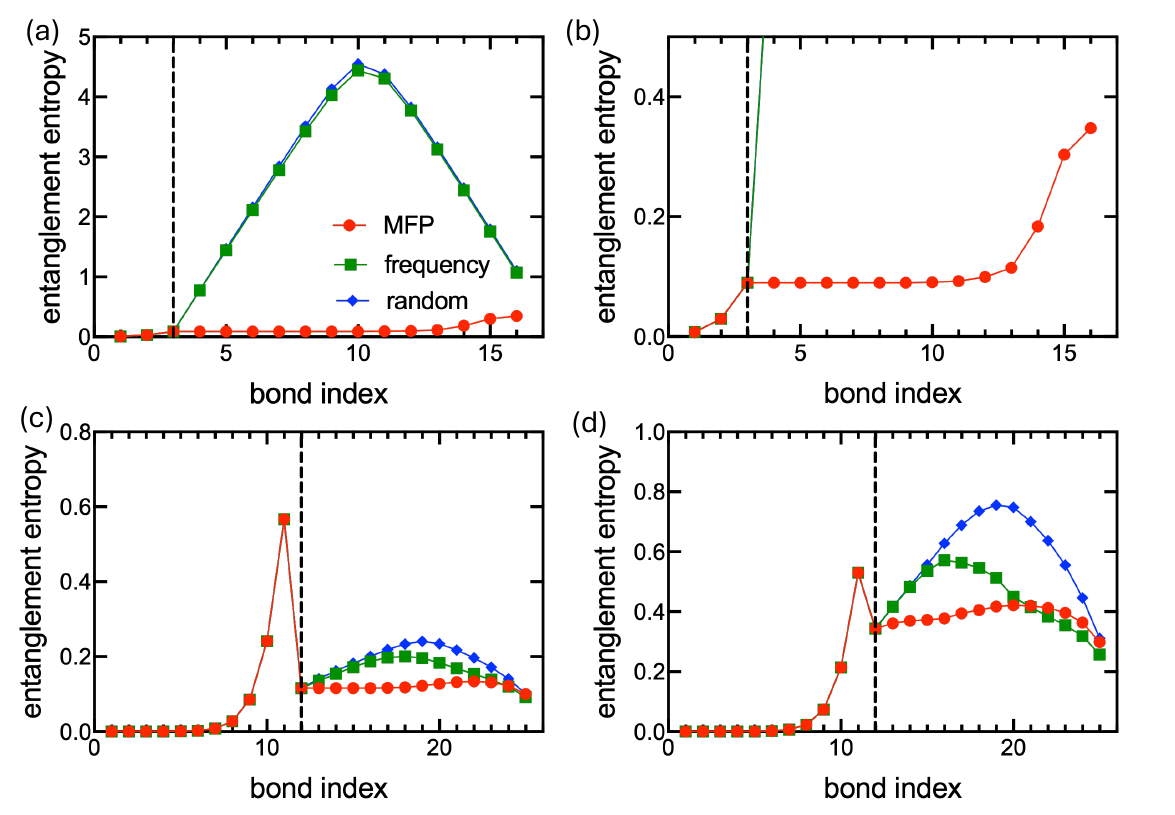}
    \caption{\label{fig:entropy}
    Entanglement entropy of MPS constructed with different indexing schemes of the modal space, compared for three different transport regimes: (a) ballistic ($L=1$ nm), (b) the same ballistic case with a smaller scale on the y-axis, (c) quasi-ballistic ($L=$1 $\mu$m), and (d) diffusive ($L \to \infty$). The vertical dash line represents the bond connecting tensors associated with real and modal spaces.}
\end{figure}

The real-space domain is discretized with $2^3$ and $2^{12}$ control volumes for $L=1$~nm and $L=1~\mu$m, respectively, requiring 3 and 12 qubits for the real-space portion of the MPS. The modal space is sampled on a $16 \times 16 \times 16$ wavevector grid with 6 phonon branches, requiring 13 qubits and 1 qutrit ($d=3$). The qutrit is placed at the center of the modal-space block.

\begin{figure}[t]
    \vspace{0.1cm}
    \centering
    \includegraphics[width=0.95\linewidth]{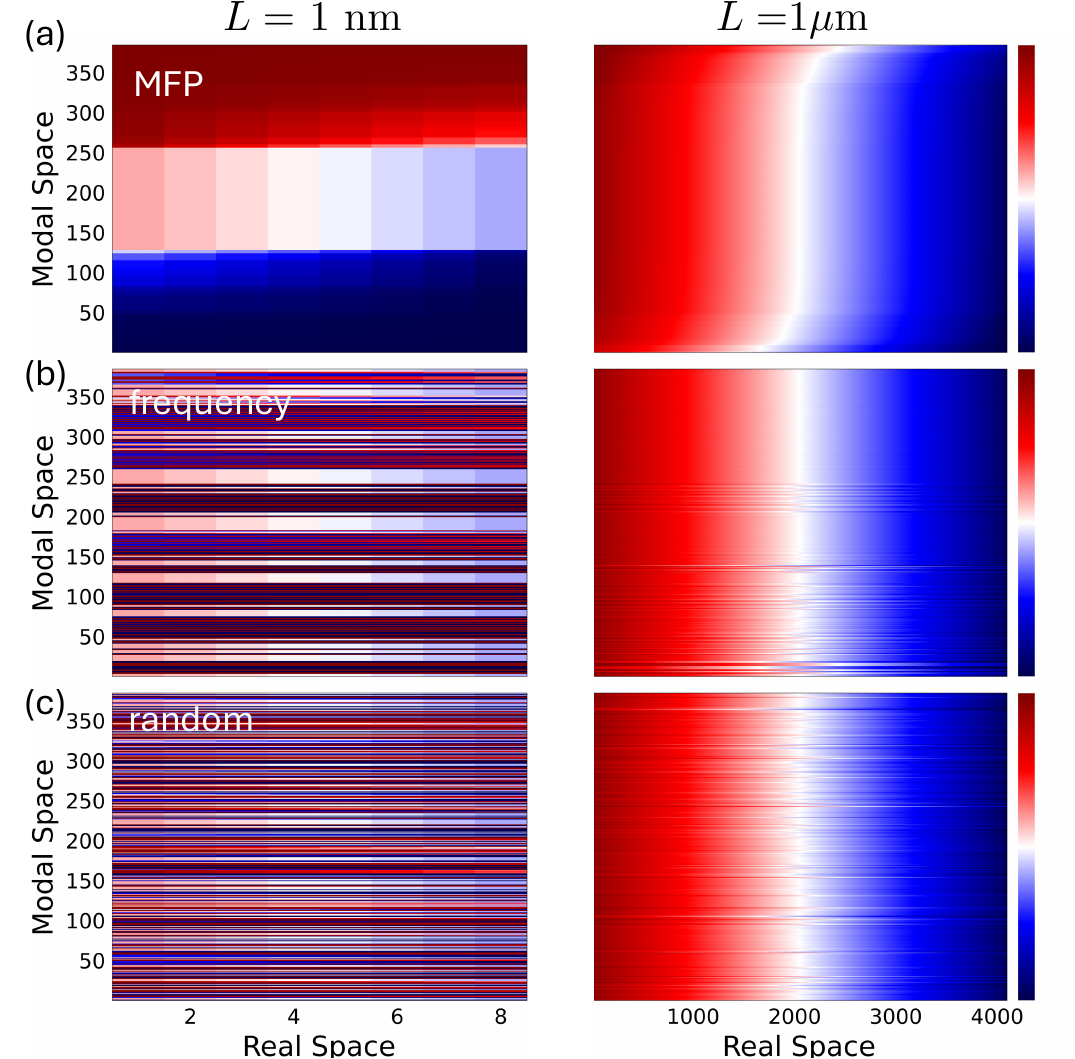}
    \caption{\label{fig:FVMsol}
    The distribution $f^*$ on the plane of modal and real space indices from the FVM solution. The modes are indexed based on (a) MFP, (b) frequency, and (c) randomly. In (a), the mode indices are assigned in ascending order of MFP from negative (propagating to $-x$) to positive (propagating to $+x$). In (b), the mode indices are assigned in ascending order of frequency from the lowest to the highest. For the purpose of visualization, a coarse grid ($4 \times 4\times 4$) is used for the wavevectors with 6 branches. The fully diffusive case ($L \to \infty$) is qualitatively similar to the quasi-ballistic case ($L=1$ $\mu$m). The colors represent values of $f^*$ ranging from 0.5 (dark red) to -0.5 (dark blue).
    }
\end{figure}

Figure~\ref{fig:entropy} shows the entanglement entropy of the MPS under the three indexing schemes for different transport regimes. We first compare the entanglement entropy in the real space portion of the MPS. While this entropy is small for the ballistic case, it is noticeably large at bonds connecting coarse real space tensors in the quasi-ballistic and diffusive cases. To provide physical insight, we show the distribution $f^*$ on the phase space in Fig.~\ref{fig:FVMsol}. In the ballistic case, the majority of modes have a nearly constant distribution in real space, retaining the distribution with which they were emitted from the left and right boundaries. As the distribution is a nearly constant function in real space, it can be encoded into the MPS with little entanglement between subsystems. However, in the quasi-ballistic and diffusive cases, the distribution gradually varies in real space, leading to a large entanglement entropy at the bonds connecting the coarse real space tensors.

\begin{figure}[t]
    \vspace{0.1cm}
    \centering
    \includegraphics[width=1.0\linewidth]{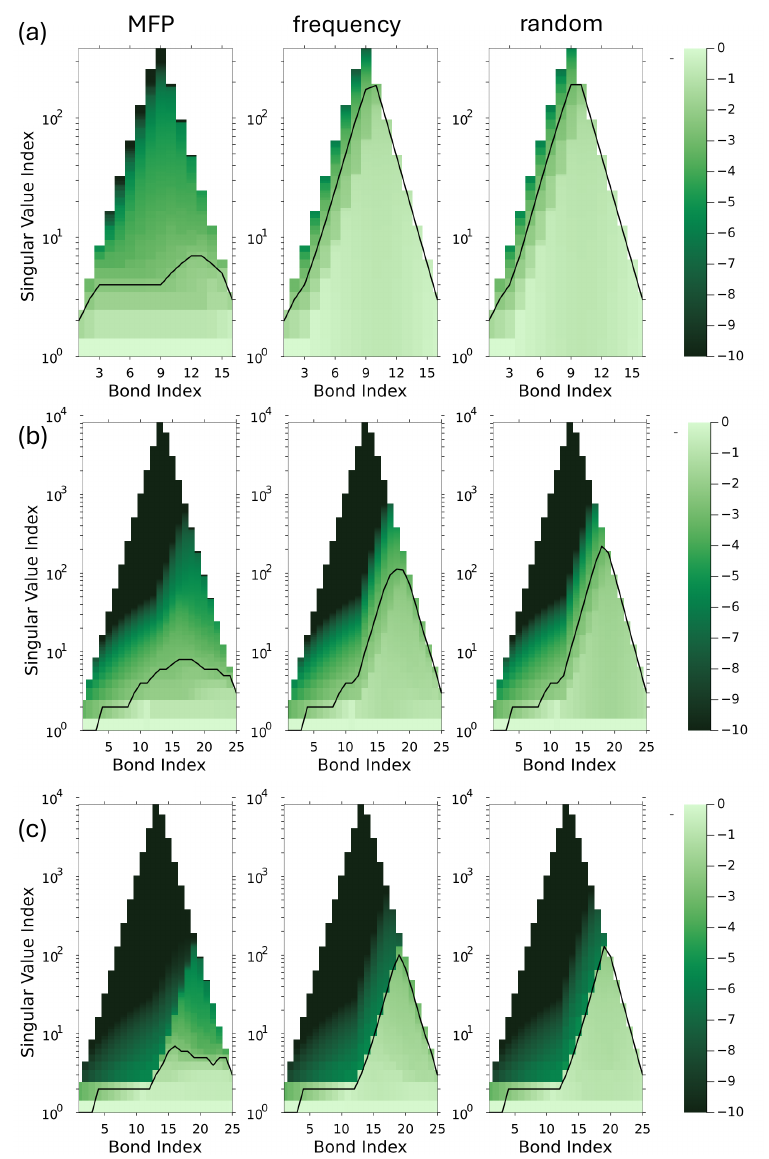}
    \caption{\label{fig:schmidt}
    Schmidt spectrum of MPS constructed with different indexing schemes for the modal space, compared for three different transport regimes: (a) ballistic ($L=$1 nm), (b) quasi-ballistic ($L=$1 $\mu$m), and (c) diffusive ($L \to \infty$). The color bar on the right shows $\log(\lambda)$ where $\lambda$ is the Schmidt value. The black line represents $\chi_l$ satisfying $\epsilon_l(\chi_l) < 10^{-5}$.
    }
\end{figure}

The entanglement entropy in Fig.~\ref{fig:entropy} also strongly depends on the indexing scheme for the modes. In all transport regimes, the frequency- and random-indexing schemes show substantially larger entanglement entropy than the MFP-indexing scheme. This trend is much more notable in the ballistic case. Figure~\ref{fig:FVMsol} clearly illustrates the underlying reason. When the MFP-indexing scheme is employed, the distribution is a relatively smooth and monotonous function in the modal space for all transport regimes; thus the tensors associated with different MFP-scales are weakly correlated. However, when frequency- and random-indexing schemes are used, the distribution in the ballistic regime is a rapidly fluctuating function in the modal space, requiring all the tensors across different scales to be strongly correlated. In the quasi-ballistic and diffusive regimes, more frequent scattering events drive the distribution closer to local equilibrium, resulting in a smoother function in the modal space while still retaining some degree of fluctuation. As a result, the entanglement entropy in the modal space for quasi-ballistic and diffusive regimes is smaller than that for the ballistic regime.

We further examine the degree of locality in the different indexing schemes using the Schmidt spectrum in Fig.~\ref{fig:schmidt}. The Schmidt spectrum provides the full details of the decay of singular values at each bond, which cannot be fully expressed by the value of the entropy. We also show the number of singular values, i.e., the bond dimension $\chi_l$, required for a truncation error per bond below $10^{-5}$ at bond $l$. This is indicated by black solid lines in each panel. The truncation error at bond $l$ upon retaining only the $\chi_l$ largest singular values after each SVD is found based on the discarded singular values as shown in Eq.~\eqref{eq:epsilon_l}. The total error of the truncated MPS, $\ket{f^*_\chi}$, relative to the untruncated MPS, $\ket{f^*}$, defined as $\epsilon_T = \norm{\ket{f^*_\chi}-\ket{f^*}} / \norm{\ket{f^*}}$, is bounded as~\cite{verstraeteMatrixProductStates2006}
\begin{equation}
    \label{eq:epsilon_T}
    \epsilon_T \leq \sqrt{2 \sum_{l=1}^{N-1} \epsilon_l(\chi_l)}.
\end{equation}

Thus, truncating bonds with $\epsilon_l$ of $10^{-5}$ roughly yields a total error of order $10^{-2}$ given that our MPS has 16 to 25 bonds. In other words, the black solid line in Fig.~\ref{fig:schmidt} represents the required bond dimension ($\chi_l$) at each bond to compress the distribution $f^*$ with a total error $\epsilon_T$ below $10^{-2}$. In the ballistic regime, the Schmidt spectrum decays so slowly under the frequency- and random-indexing schemes that almost every singular value needs to be included. The Schmidt spectrum in the quasi-ballistic regime shows a much faster decay of singular values compared to the ballistic case. A bond dimension of less than 10 is sufficient for the MFP-indexing scheme, while bond dimension above 100 is required for the frequency- and random-indexing schemes. In the diffusive regime, the Schmidt spectrum shows the fastest decay of singular values among the three transport regimes for both real and modal spaces. A bond dimension of $\chi_l=2$ is sufficient for the real space, indicating the potential for significant compressibility of the data.

\begin{figure}[b!]
\vspace{0.1cm}
\centering
\includegraphics[width=0.95\linewidth]{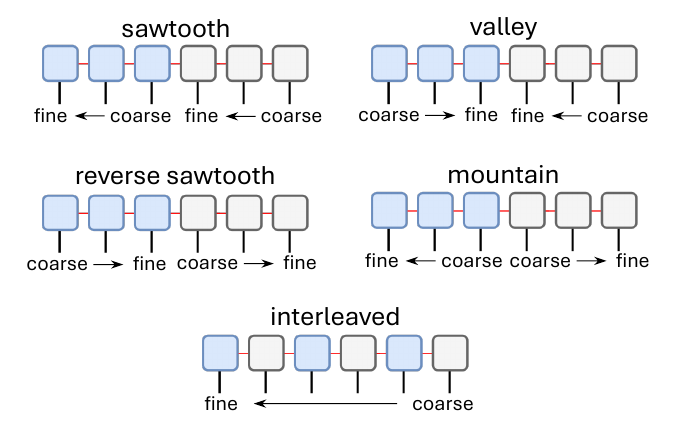}
\caption{\label{fig:MPSconfig2}
Various MPS configurations. Blue and grey tensors are associated with the real and modal spaces, respectively.
}
\end{figure}

\begin{figure*}[t]
    \vspace{0.1cm}
    \centering
    \includegraphics[width=0.95\linewidth]{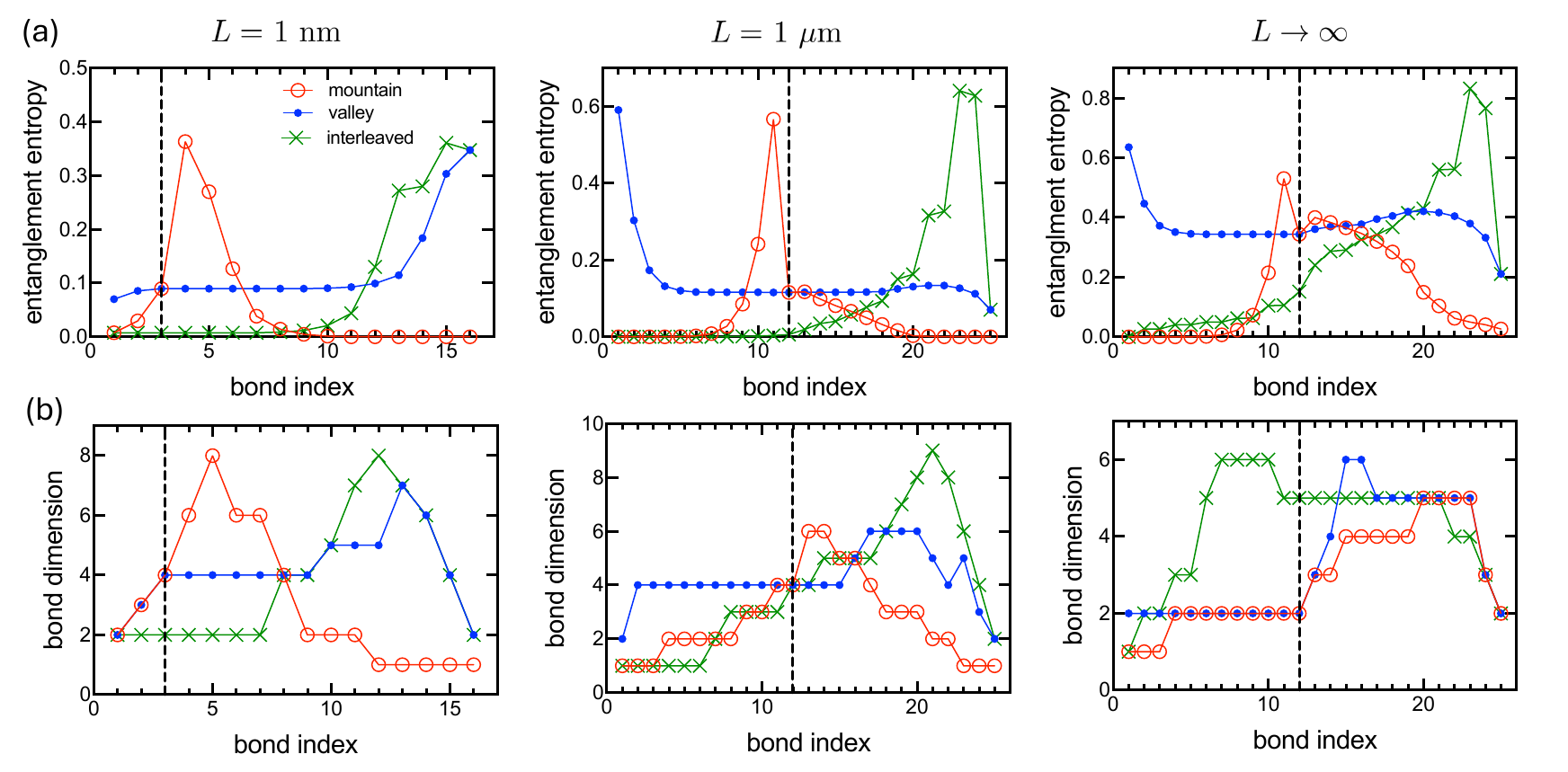}
    \caption{\label{fig:entropy_bonddim_v2}
    Comparison of various MPS configurations in terms of (a) entanglement entropy and (b) bond dimension ($\chi_l$) required to achieve $\epsilon_l(\chi_l) < 10^{-5}$. Bonds to the left and right of the vertical line correspond to the real- and modal-space partitions, respectively, except in the interleaved configuration. The sawtooth configuration combines mountain ordering for real space and valley ordering for modal space; the reverse sawtooth adopts the opposite arrangement.
}
\end{figure*}

We further explore various MPS configurations using the MFP-indexing scheme for the modal space. Fig.~\ref{fig:MPSconfig2} shows the five MPS configurations examined in this work. In the first four configurations (sawtooth, valley, reverse sawtooth, and mountain), the real-space and modal-space tensors are separated into distinct blocks, but the tensors within each block are arranged in different orders of significance. The sawtooth configuration is used for Fig.~\ref{fig:entropy} and Fig.~\ref{fig:schmidt}. We also examine an interleaved configuration in which the tensors for real space and modal space alternate along the MPS chain, pairing tensors associated with similar length scales and MFP scales on neighboring sites. As we have more tensors for the modal space than for the real space, i.e., $N_i>N_x$, we place all the modal space tensors that are not paired with a real space tensor in the middle of the chain. For example, $N_x$ and $N_i$ are 12 and 14, respectively, for the quasi-ballistic and diffusive regimes. In these cases, an MPS consists of 6 pairs of alternating real and modal space tensors representing fine grid, 2 unpaired modal space tensors, and 6 pairs of alternating real and modal space tensors representing coarse grid from left to right ends of the chain.

Fig.~\ref{fig:entropy_bonddim_v2} shows the entanglement entropy and the required bond dimension ($\chi_l$) for a truncation error threshold of $\epsilon_l = 10^{-5}$ when the reference solution $\mathbf{f}^*_\text{ref}$ is encoded into an MPS under each configuration. For the ballistic case, the bond dimensions within the real-space portion of the MPS are uniformly small across all configurations. This is expected because, as shown in Fig.~\ref{fig:FVMsol}, the distribution function is nearly constant in real space. Phonons traverse the domain with negligible scattering, so the spatial profile carries little information regardless of how the real-space tensors are ordered. In contrast, the modal-space portion reveals clear differences among configurations. The distribution function varies significantly over the large MFP scale, meaning that the coarsest modal-space tensors, those distinguishing modes with vastly different MFP, carry the most information. In the valley and sawtooth configurations, these coarse grid tensors are placed at the outer edge of the modal-space block, far from the junction with the real-space tensors. As a result, the information encoded in the coarse tensors must propagate through the entire chain of modal-space sites to reach the rest of the MPS, leading to elevated entanglement entropy and large $\chi_l$ throughout the modal-space block. In the mountain and reverse sawtooth configurations, the coarse modal-space tensors are instead positioned at the junction with the real-space block, so the most important information is already located near the boundary between the two spaces and does not need to traverse the full modal-space chain. This significantly reduces both the entropy and the required $\chi_l$.

The entropy and required $\chi_l$ in the quasi-ballistic and diffusive regimes can be understood similarly. In these cases, unlike the ballistic regime, the distribution function also varies appreciably in real space, predominantly over the large length scale. This makes the mountain configuration preferable to the reverse sawtooth configuration for the real space block, as the coarsest real-space tensor is placed at the junction rather than at the outer edge.

\begin{figure*}[t]
    \vspace{0.1cm}
    \centering
    \includegraphics[width=0.95\linewidth]{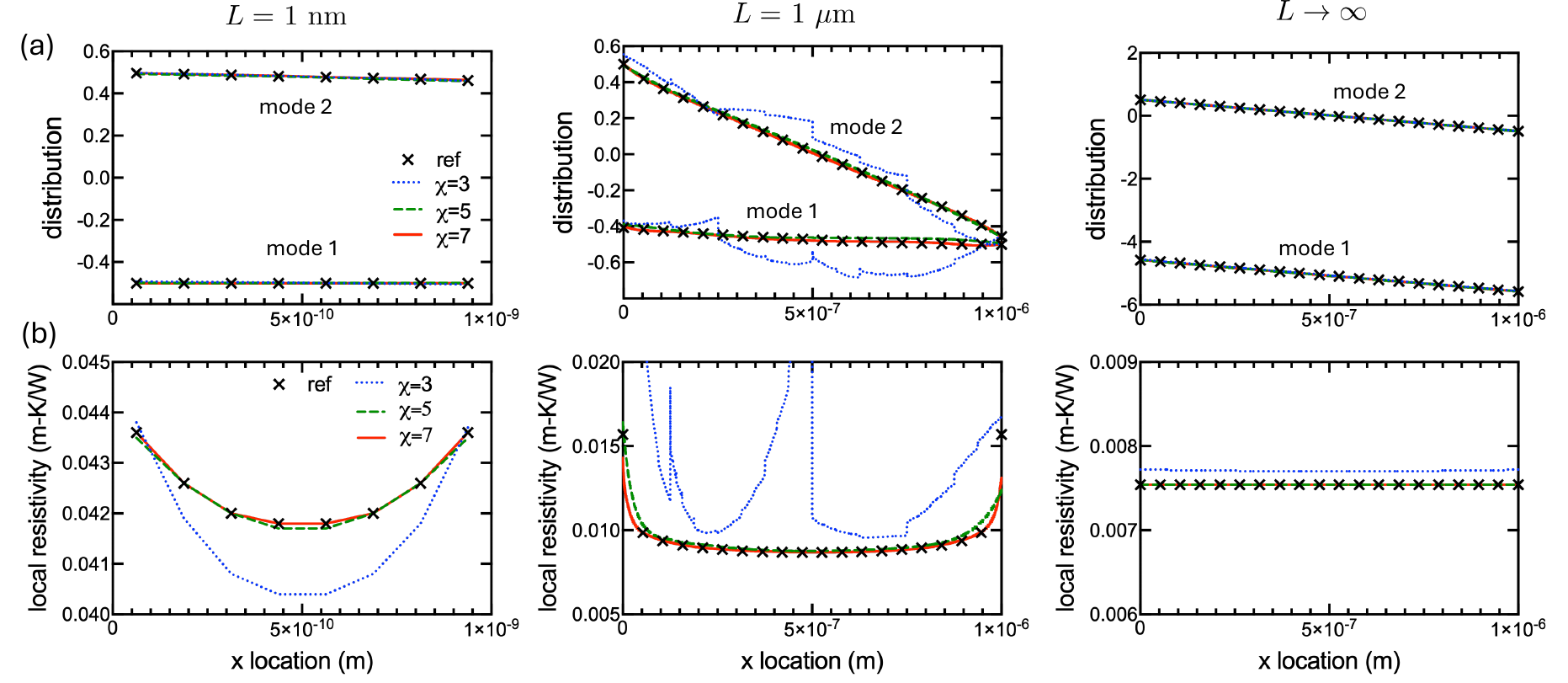}
    \caption{\label{fig:res_dist}
    TNFVM results for ballistic, quasi-ballistic, and diffusive cases with different $\chi_{\max}$ values: (a) distribution $f^*$ for the two representative LA modes and (b) local resistivity.
    }
\end{figure*}

Across all three transport regimes, the mountain configuration performs best overall. It places the coarsest---and most information-rich---tensors of both the real-space and modal-space blocks at the center of the MPS chain, at the junction between the two blocks. This arrangement minimizes the distance over which information must flow through the chain, thereby minimizing the entanglement entropy and the required bond dimension. The required bond dimension for the mountain configuration ranges from 5 to 8 depending on the transport regime, representing a highly compact MPS representation. The interleaved configuration, despite its intuitive appeal of pairing tensors at similar scales, does not outperform the mountain configuration in our case.

\subsection{TNFVM solution}
We solve the PBE using the TNFVM described by Eq.~\eqref{eq:TNFVM} and the DMRG-like method of Eq.~\eqref{eq:DMRG} under different truncation conditions. In all calculations, the truncation threshold is set to $\epsilon_l = 10^{-5}$, and the maximum bond dimensions across the entire MPS ($\chi_{\max}$) are bounded to 3, 5, and 7.

To illustrate the accuracy of the MPS-based solutions, Fig.~\ref{fig:res_dist}(a) presents $f^*$ for two representative phonon modes that exhibit contrasting transport behavior. Mode~1 is a low-frequency longitudinal acoustic (LA) mode with a frequency of 1.89~THz and an MFP of $-5.08~\mu$m where the negative sign indicates the mode is propagating along $-x$ direction. Thus, the mode undergoes nearly ballistic transport across the $1~\mu$m sample. Mode~2 is a high-frequency LA mode with an MFP of 12.3~nm and a frequency of 12.0~THz, which experiences nearly diffusive transport in the same sample. For both modes and across all three transport regimes, the solutions obtained with $\chi_{\max} = 5$ and 7 are nearly indistinguishable from the reference solution, $\mathbf{f}^*_\mathrm{ref}$. In the ballistic and diffusive cases, even a very small bond dimension of $\chi_{\max} = 3$ is sufficient to reproduce $\mathbf{f}^*_\mathrm{ref}$ with high fidelity. In the quasi-ballistic case, however, $\chi_{\max} = 3$ produces noticeable deviations from the reference solution.

Figure~\ref{fig:res_dist}(b) presents the local resistivity, $\rho(x)$, calculated as~\cite{ziman_electrons_1960, li_thermal_2023}
\begin{equation}
    \label{eq:resistivity}
    \rho(x) = \left( \frac{T(x)}{q''} \right)^{2} \frac{k_B}{N_q V_\text{uc}} \sum_{i=1}^{n_i} \frac{\left( f_{x,i} - f_{x,i}^{0} \right)^{2}}{f_{x,i}^{0} \left( f_{x,i}^{0} + 1 \right) \tau_i},
\end{equation}
where $q'', k_B, N_q, V_\text{uc}$ are heat flux, the Boltzmann constant, number of wavevectors considered, and the volume of unit cell. The local resistivity quantifies the variance of the non-equilibrium distribution function relative to the local equilibrium distribution. Because the local resistivity involves a sum of squared deviations over all modes, errors from individual modes accumulate constructively rather than canceling, making it a sensitive measure of the overall solution accuracy.

In all three transport regimes, $\chi_{\max} = 3$ shows noticeable deviations in the local resistivity profile from the reference solution. The error is particularly pronounced in the quasi-ballistic case. As $\chi_{\max}$ increases, the MPS solutions converge systematically toward the reference data. At $\chi_{\max} = 7$, the local resistivity profiles are nearly identical to the reference solution across all transport regimes, confirming that a bond dimension of 5 to 7 is sufficient to accureately represent the phonon distribution function in the mountain MPS configuration across the range of transport conditions considered here.

\subsection{Computational cost}
We compare the computational cost of the conventional FVM and the TNFVM for the quasi-ballistic regime. Both methods are run to achieve a relative error $\lVert \mathbf{f}^* - \mathbf{f}^*_{\text{ref}} \rVert / \lVert \mathbf{f}^*_{\text{ref}} \rVert < 10^{-2}$. The TNFVM calculations use $\chi_{\max} = 7$ and $\epsilon_l = 10^{-5}$. The reported CPU time accounts only for the solver itself. For the FVM, this corresponds to the time spent on solving Eq.~\eqref{eq:FVMitercompact}, while for the TNFVM, it corresponds to the time spent in the DMRG-like iterative solver following Eq.~\eqref{eq:TNFVM} and Eq.~\eqref{eq:DMRG}. The computational cost of pre- and post-processing is excluded from the CPU time reported in this work. The TNFVM requires a pre-processing step in which phonon properties such as MFP and scattering rates are encoded in MPS form. However, this encoding needs to be performed only once for each material, and its computational cost is much lower than that of calculating MFP and scattering rates from first principles. Both the FVM and TNFVM require post-processing to calculate quantities such as local temperature, heat flux, and thermal resistivity. In the TNFVM, this post-processing can be performed efficiently by evaluating inner products and expectation values directly in MPS form, avoiding the need to contract the MPS into a full array. All calculations are performed on a single core of an AMD EPYC 9374F (Genoa) processor.

\begin{figure}[htbp]
    \vspace{0.1cm}
    \centering
    \includegraphics[width=0.8\linewidth]{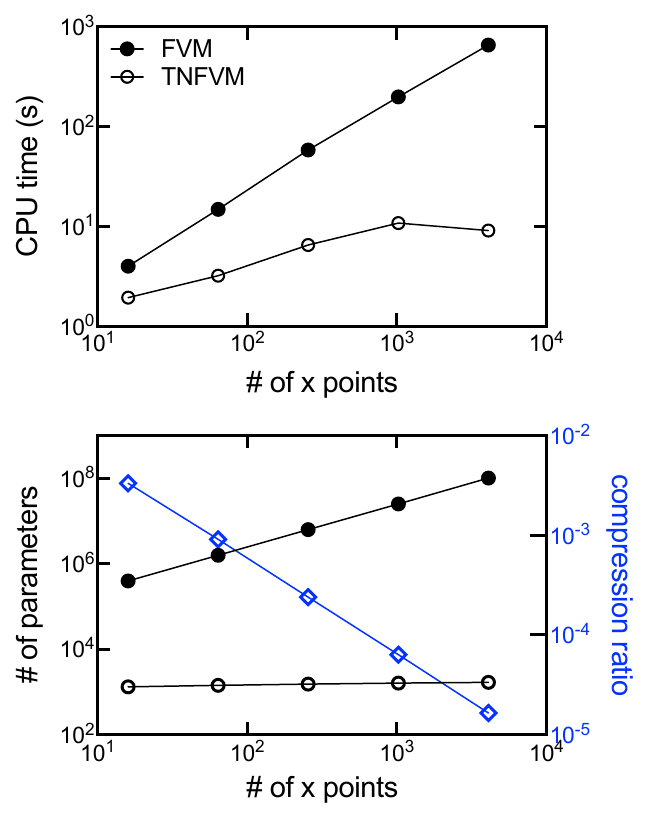}
    \caption{\label{fig:cost1}
    Computational cost with respect to the number of grid points in real space. The number of $x$ points, $n_x$, are $2^4$, $2^6$, $2^8$, $2^{10}$, and $2^{12}$.
    }
\end{figure}

We first examine how the computational cost scales with the number of real-space grid points in Fig.~\ref{fig:cost1}. The modal space is sampled on a $16 \times 16 \times 16$ wavevector grid with 6 branches, which is known to be reasonably sufficient to converge the bulk thermal conductivity. We note that a real-space grid of $2^8$ control volumes, corresponding to a spatial resolution of 3.9~nm, is already sufficient to converge the quasi-ballistic simulation; finer grids up to $2^{12}$ are examined here solely to probe the scaling behavior of each method.

\begin{figure}[htbp]
    \vspace{0.1cm}
    \centering
    \includegraphics[width=0.8\linewidth]{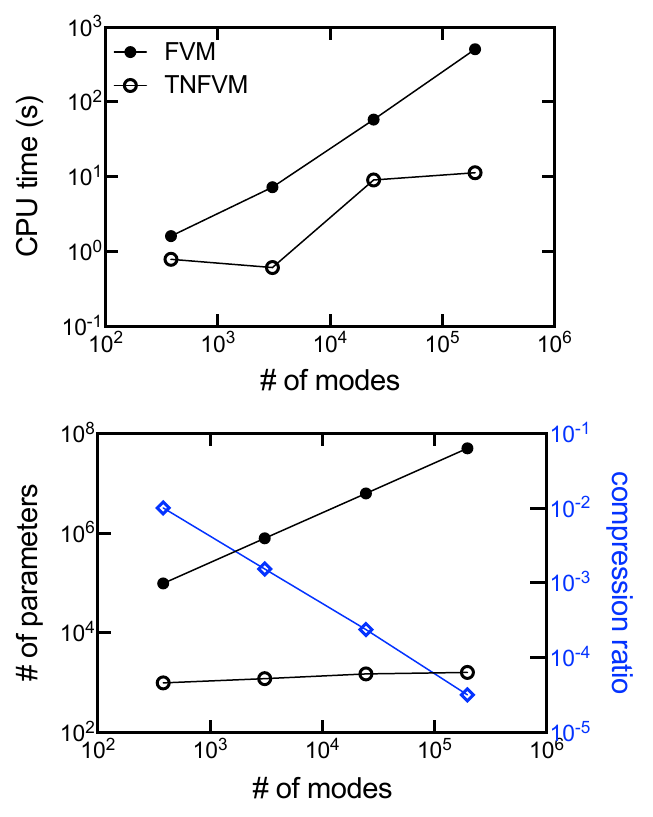}
    \caption{\label{fig:cost2}
    Computational cost with respect to the number of grid points in modal space, $n_i$. The number of wavevector points are $4^3$, $8^3$, $16^3$, and $32^3$ with 6 branches.
    }
\end{figure}

For the FVM, the CPU time scales linearly with the number of real-space grid points, as expected from the treatment of $\mathbf{H}$ as a sparse matrix whose size grows proportionally with the number of spatial degrees of freedom. In contrast, the TNFVM exhibits sublinear scaling; the CPU time increases more slowly as the real-space grid is refined, and notably does not increase further when the number of grid points grows from $2^{10}$ to $2^{12}$. This saturation reflects the fact that introducing a finer real-space resolution results in additional qubits whose entanglement entropy with the qubits that encode larger length scales is negligible. The fine-scale spatial variations of the distribution function are small, and thus the DMRG-like solver converges with minimal additional effort. At $2^8$ real-space grid points, the TNFVM already exhibits roughly one order of magnitude lower CPU time compared to the conventional FVM.

To further quantify the efficiency of the MPS representation, we compare the number of parameters required by each method. For the FVM, the total number of parameters is $n_{\text{FVM}} = n_xn_i$. For the TNFVM, the number of parameters in the MPS is $n_{\text{TNFVM}} = \sum_{j=1}^{N_x+N_i} d_j \, \chi_{j-1} \, \chi_j$, where $d_j$ is the dimension of the physical site $j$ (namely, $d_j=2$ for qubits and $d_j=3$ for qutrits). We define the memory compression ratio as $n_{\text{TNFVM}} / n_{\text{FVM}}$. The number of parameters for the FVM increases linearly with the number of real-space grid points, as the full solution vector grows proportionally with the grid size. In contrast, the number of parameters in the TNFVM barely increases as the real-space grid is refined. Each additional real-space qubit appended to the MPS introduces only a small tensor whose size is bounded by $\chi_{\max}$, which remains nearly unchanged because the fine-scale spatial structure of the distribution function carries little additional information. This results in a compression ratio that decreases rapidly with grid refinement, demonstrating the significant compressibility afforded by the MPS ansatz.

We next examine how the computational cost scales with the number of phonon modes, fixing the real-space grid at $2^8$ control volumes. The results are shown in Fig.~\ref{fig:cost2}. For the FVM, both the CPU time and the number of parameters scale linearly with the number of modes. For the TNFVM, both quantities scale sublinearly with the number of modes. As additional modal-space qubits are appended to the MPS, they introduce tensors whose bond dimensions remain bounded by $\chi_{\max}$, resulting in a cost that grows much more slowly than the full grid size. Notably, the CPU time slightly decreases between the first two points in Fig.~\ref{fig:cost2}(a). In both cases, the DMRG-like solver converges rapidly owing to the small problem size. The first case already exhibits small error close to the convergence criteria after 2 sweeps but requires third sweep to meet the criterion $\lVert \mathbf{f}^* - \mathbf{f}^*_{\text{ref}} \rVert / \lVert \mathbf{f}^*_{\text{ref}} \rVert < 10^{-2}$. The second case marginally meets the criterion after only 2 sweeps, accounting for the slight reduction in CPU time. Furthermore, a very large memory compression due to the MPS representation is found, similar to that shown in Fig.~\ref{fig:cost1}. As a consequence, the computational cost of the TNFVM remains low even when a very fine modal-space grid is employed. This feature is particularly advantageous in situations where a fine wavevector grid is physically necessary, for example when heat is carried by the modes near the Brillouin zone center only and thus require dense sampling of the wavevector space. Such conditions arise in materials like diamond at room temperature or in silicon at cryogenic temperatures.

\section{Conclusion}

We explored the use of MPS for solving the PBE with full-dimensional modal space and \textit{ab initio} phonon properties, addressing the curse of dimensionality inherent in the PBE. Significant correlation locality of the solution MPS is observed when the modal space is indexed based on MFP and the distribution is expressed in dimensionless form. This can be understood from the fact that the distribution function of each mode is coupled to those of other modes through scattering processes. Roughly speaking, phonon modes that strongly interact with one another share similar temperatures, while phonon modes that rarely interact with others have more independent distribution functions. Expressing the distribution in dimensionless form preserves this similarity across a wide range of the phonon spectrum, and decomposing the solution tensor into tensors associated with different length scales of MFP leads to strong correlation locality in the MPS. We further investigated the ordering of tensors within the MPS chain. The highest correlation locality is observed in the mountain configuration, where the tensors associated with the coarsest grids for real-space position and MFP are placed together at the center of the MPS chain. This configuration minimizes the distance over which information must propagate along the chain, thereby minimizing the entanglement entropy and the required bond dimension at every bipartition.

Using the optimal MPS configuration, we solved the PBE with a DMRG-like method across the ballistic, quasi-ballistic, and diffusive transport regimes. A bond dimension of 5 to 7 is sufficient to reproduce the reference FVM solution with high fidelity for both the mode-resolved distribution function and the local thermal resistivity. The computational cost of the TNFVM scales sublinearly with the number of grid points in both real and modal spaces, in contrast to the linear scaling of the conventional FVM. This sublinear scaling originates from the fact that additional qubits appended to the MPS introduce tensors with negligible entanglement, leaving the solver effort nearly unchanged. As a result, the TN approach achieves roughly one order of magnitude reduction in CPU time and three orders of magnitude reduction in memory at moderate grid sizes and offers increasingly favorable compression ratios as the grid is refined.

These findings demonstrate that the MPS-based approach provides an efficient and accurate framework for solving the PBE with the full \textit{ab initio} phonon dispersion and scattering rates. Future work may extend this approach to higher-dimensional real spaces and multi-carrier transport, where the advantages of TN methods in managing the curse of dimensionality are expected to be even more pronounced.

\begin{acknowledgments}
S.L. acknowledges support from the National Science Foundation (Award No. 2518562). This research was also supported in part by the University of Pittsburgh Center for Research Computing, RRID:SCR022735, through the resources provided. Speciﬁcally, this work used the H2P cluster, which is supported by NSF Award No. OAC-2117681. S.L. thanks Nikita Gourianov for the help in calculating entanglement entropy and Schmidt spectrum.
\end{acknowledgments}

\bibliography{references}

\end{document}